# FREE-SPACE GRAPHENE/SILICON PHOTODETECTORS OPERATING AT 2 MICRON

M. Casalino,[1] R. Russo,[1] C. Russo,[2] A. Ciajolo,[2] E. Di Gennaro,[3] M. Iodice,[1] G. Coppola[1]

[1]*Istituto per la Microelettronica e Microsistemi IMM-CNR – Via P. Castellino n. 111- 80131 Napoli (Italy)*
[2]*Istituto di Ricerche sulla Combustione IRC-CNR - P.le V. Tecchio n. 80 - 80125 Napoli (Italy)*
[3]*Dip. di Fisica "E. Pancini", Universitá di Napoli Federico II and CNR-SPIN Compl. Univ. di Monte S. Angelo, Via Cintia, I-80126 Napoli, Italy*

**This paper presents the design, the fabrication and the characterization of Schottky graphene/silicon photodetectors, operating at both 2 μm and room temperature. The graphene/silicon junction has been carefully characterized: device shows a non–ideal behaviour with the increasing temperature and the interfacial trap density has been measured as $1.1 \times 10^{14}$ eV$^{-1}$cm$^{-2}$. Photodetectors are characterized by an internal (external) responsivity of 10.3 mA/W (0.16 mA/W) in an excellent agreement with the theory. Our devices pave the way for developing hybrid graphene-Si free-space illuminated PDs operating at 2 μm, for free-space optical communications, optical coherence tomography and light-radars.**

## 1. Introduction

Recently, many research groups have demonstrated several silicon(Si)-based components operating in the mid-wave infrared (MIR) wavelength range of 2–20 μm, including low-loss waveguides, couplers, splitters and multiplexers [1], as well as some with hybrid active functionality [2, 3]. There are compelling reasons to migrate Si photonics from the telecom wavelength region into the MIR. First, the undesired nonlinear loss, two-photon absorption (TPA), which is a limiting factor for nonlinear optical processes in the near-infrared vanishes at longer wavelengths as the energy of two photons is not enough for a band-to-band transition [4]. Second, Si photonics have many potential applications in chemical and biological sensing for realizing the lab-on-a-chip concept. Low cost miniature Si sensors for trace gas detection, bio-agent sensing, environmental monitoring and industrial process control will attract researchers' interest, too. In particular, the wavelength of 2 micron results very interesting for different reasons, among them the possibility: to confine the light by using silicon-on-insulator (SOI) waveguides (at longer wavelengths the buried oxide greatly absorbs the propagating modes), to amplify the light by the stimulated Raman scattering (SRS) due to the reduced TPA absorption and to detect gas like carbon dioxide ($CO_2$) characterized by an absorption window close to 2 μm. For all the aforementioned applications, the photodetectors (PDs) are key components. Although Si optical detectors are widely used for visible light (400-700nm), unfortunately, they can not work at wavelengths longer than ~1.1μm due to the Si bandgap of 1.12eV. However, extensive efforts have been employed in order to realize Si-based PDs at Near-Infraed (NIR) regime for telecommunications [5-11]. In this context, the significant success of the integration of germanium with silicon to produce fast photodetection at wavelengths around 1.3 and 1.55 μm [12, 13] cannot be extended to 2 μm because of the cut-off wavelength of germanium [14]. There has been recent interest in expanding the silicon photonics platform to function at longer wavelengths in the region of 2 μm [15]. Indeed, all-Si photodiodes can provide sensitivity that extends into this wavelength range taking advantage of defect-mediated absorption, i.e., lattice defects voluntary introduced into Si associated to deep-level charge states within the Si bandgap, responsible of the displacement of the Si optical absorption curve towards longer wavelengths [16]. Performance at 1 Gb/s was demonstrated by Souhan *et al.* for Si+ implanted P-I-Ns to demonstrate 10 mA/W sensitivity at 2.2 μm wavelength [17]. Similarly, B+ implanted large cross-section waveguides were tested across 2-2.5 μm wavelength and the results show responsivity of 3 mA/W at 2 μm [18]. Ackert *et al.* designed a defect-mediated APD with waveguide dimensions suitable for 2 μm wavelength operation [19]. In this case the implanted B+ ions allow achieving a responsivity of 0.3 A/W for 2.02 μm wavelength on a 0.2 mm long device. These results confirm both the suitability of defect-mediated detectors for MIR wavelengths but they are always realized by waveguide structure, mainly to take advantage of the long light-matter interaction. However, to the best of our knowledge, free-space Si PDs have not been reported as far even if they could be very useful for applications where waveguiding structure can not be adopted. Indeed the wavelength of 2 μm could be used in free-space optical communications (FSO) [20, 21] and light-radars (LIDARs) [22, 23] applications where the propagation losses in fog[12] and humid conditions[15] are minimized due to lower optical absorption and scattering compared to wavelengths < 1μm. In addition the eye safety is improved because the outer layer of the eye (cornea) absorbs light at 2 μm and does not allow it to focus on retina[12, 13]. In addition, optical coherence tomography (OCT) is a non-invasive imaging technique for biological tissues[16] where the advantages of using 2 μm are, the enhanced penetration depth, due to lower scattering in tissue with respect to shorter wavelengths[16], and the enhanced imaging contrast at deeper penetration depths, where multi-scattering processes dominate[16].

An interesting alternative approach, with potential for integration with silicon, is to take advantage of the internal photoemission effect (IPE) [26-29]. IPE is the optical excitation of the electrons in a metal to energy above the Schottky barrier and the corresponding transport of these electrons to the semiconductor conduction band [30-34]. Although Schottky PDs are typically realized with metals, very recently graphene has shown the capability to enhance IPE [35]. Moreover, hybrid graphene/Si structures have already shown their potentialities also in the field of light emission [36] and modulation [37].

In this work we report on vertically-illuminated graphene/silicon PDs based on IPE operating at 2 μm. The graphene/silicon junction has been carefully investigated by temperature dependent I-V measurements between 280-315 K. Non-ideal behavior such as increase in zero bias Schottky barrier height (SBH) and decrease in ideality factor with increasing temperature has been observed. This behavior has been ascribed to spatial fluctuations of the built-in voltage $V_{bi}$ and consequently of SBH inhomogeneities. In addition, the interface quality has been investigated by measuring the interface trap density that results as high as $1.1 \times 10^{14}$ $eV^{-1}cm^{-2}$. Finally, free-space graphene/silicon PDs show an internal (external) responsivity of about 10 mA/W (0.16 mA/W) at 2 μm, room temperature and without any bias voltage applied. Our results demonstrate the graphene/Si Schottky PD provide an attractive solution for Si optical detection at wavelengths of 2 μm that can be useful for FSO communications, OCT and LIDAR applications.

## 2. Theoretical background of IPE

Detection mechanism of IPE is shown in Fig. 1.

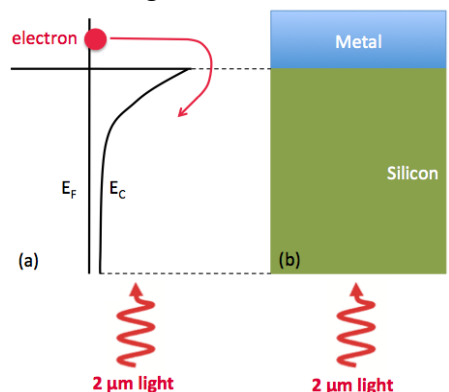

Figure 1. IPE mechanism in a metal/nSi Schottky junction: (a) energy band diagram and (b) device sketch.

IPE offers the possibility of detecting photons of energy hν lower than the Si bandgap energy provided it is greater than that of the Schottky barrier [38]. The main drawback of the Schottky PDs is their limited internal quantum efficiency $\eta_i$, defined as the number of carriers emitted to Si per absorbed photon. This is a direct result of many factors: 1) the low absorption due to high reflectivity of the metal layer, 2) the excitation of carriers lying in states far below the Fermi energy, 3) the conservation of momentum during carrier emission over the potential barrier, which lowers the carriers emission probability into a semiconductor. $\eta_i$ is linked to the PD internal responsivity $R_{int}$ defined as the ratio between the photocurrent $I_{ph}$ and the absorbed optical power $P_{abs}$, i.e $R_{int}= I_{ph}/P_{abs}= \eta_i \cdot q/h\nu$ where $P_{abs} = A\ P_{inc}$, A is the absorptance, $P_{inc}$ is the incident power, q is the electron charge. In addition, it is possible to define the external responsivity as the ratio between the photocurrent current and the incident optical power $R_{int}= I_{ph}/P_{inc}$. In classical metal/semiconductor Schottky junctions, experimental measurements are often fit to the modified Fowler equation [39]:

$$\eta_i = C \cdot \frac{(h\nu - q\phi_B)^2}{h\nu} \quad (1)$$

where C is called the quantum efficiency coefficient (C=1/8q$\Phi_B$ in agreement with the Elabd's theory [39]), hv is the photon energy, $E_F$ is the Fermi level and $\Phi_B$ the Schottky barrier.

## 3. Fabrication and Raman characterization

The device fabrication process is briefly described in this section. We start from a double-polished, low-doped (~$10^{15}$cm$^{-3}$), 200μm thick Si substrate cleaned by a standard RCA procedure [40] and dry thermal oxidised to achieve a 100nm thick $SiO_2$ layer. Subsequently, the Ohmic contact has been patterned by an optical lithography process where a positive photoresist (Shipley S1813) has been spin-coated at 4000 rpm (1.2 um thickness), UV exposed through a Ni/Cr mask and developed in order to realize a ring shape. The bare silicon dioxide has been etched in a buffer-oxide-etch (BOE) solution and without removing the photoresist, a 100nm thick Al has been thermally evaporated on the sample. Then a lift-off process and an annealing at 475 °C for 30 min is able to provide a not-rectifying behaviour contact. Two PADs have been subsequently realized by an Au layers obtained with optical lithography, Cr/Au (5nm/50nm) thermal evaporation and lift off. The Au layer protects the Al Ohmic contact form subsequent treatments involving HF. Then, the Schottky contact was realized by patterning a 50μm-diameter photoresist disk inside the Al ring, followed by $SiO_2$ etching in a BOE solution.

Finally, the sample is ready for graphene fishing. Before graphene transferring, the chip has been dipped in a mixture of hydrofluoric acid (HF) and water (2% of HF) for 1 min in order to remove the residual native oxide. Graphene was purchased from ACS Material [41]. The upper side of the graphene monolayer is covered with 500 nm of PMMA and the PMMA-graphene film is laid on a polymer substrate. To transfer the graphene, first the substrate was immersed in deionized (DI) water, which results in releasing and then floating the PMMA-graphene film on the water surface. Subsequently the PMMA-graphene film was scooped out of water using the chip. To dehydrate the sample, the graphene was blown by a nitrogen gun for a few minutes and then heated on hot plate. Finally, the sample was soaked in acetone to dissolve the PMMA and patterned by $O_2$ plasma etching carried out after a photolithographic process in order to obtain a graphene disk surrounded by an Al Ohmic ring, as shown in Fig. 2.

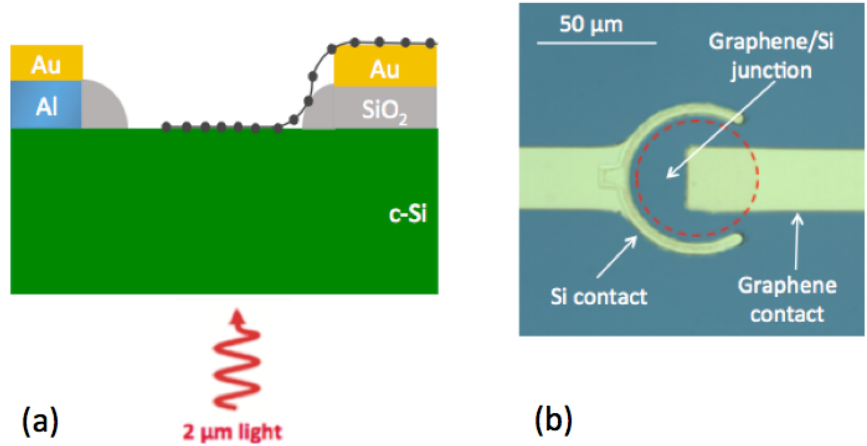

Figure 2. (a) Schematic cross-sectional view of Graphene/Si Schottky PD under illumination and (b) optical image.

On the graphene disk (red dashed circle shown in Fig. 2(b)), Raman measurements have been performed. The structural quality and doping of the graphene disk (red dashed circle shown in Fig. 2(b)) was monitored by Raman spectroscopy at 532nm using a Horiba XploRA Raman microscope system (Horiba Jobin Yvon, Japan) with a 100× objective and a laser power below 1mW. The Raman spectrum of graphene is shown in Fig.3 (black curve), where the reference Si spectrum (red curve) is also reported. The two most intense features of the graphene spectrum are the G peak and the 2D peak. The third order Si peak at ~1450cm$^{-1}$ can be also noticed.

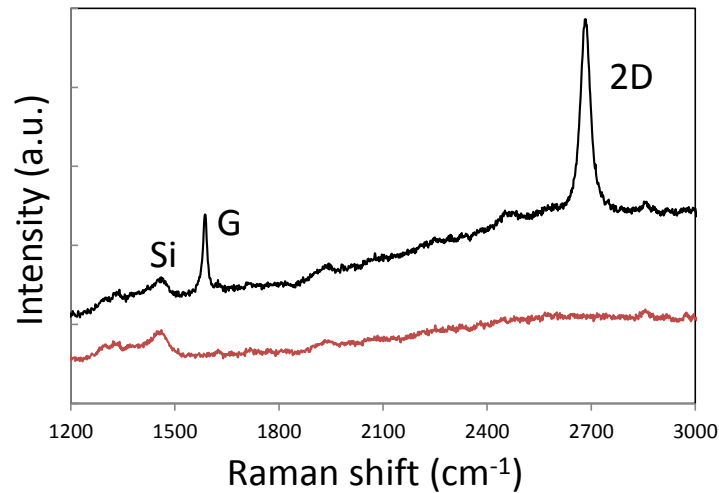

Figure 3. Raman spectra of Si substrate (red) and of graphene on Si after device fabrication (black).

The 2D peak is single-Lorentzian, signature of SLG [42]. The absence of a prominent D peak at 1350 cm$^{-1}$ indicates that there are not significant defects [42]. The position of the G peak, Pos(G), is ~1586 cm$^{-1}$ with Full Width at Half Maximum, FWHM(G), ~13.7cm$^{-1}$. The 2D peak position, Pos(2D), is~2683 cm$^{-1}$ with FWHM(2D)~29.1cm$^{-1}$. The 2D to G peak intensity and area ratios, I(2D)/I(G) and A(2D)/A(G), are ~2.5 and 5.1, respectively, suggesting a negligible doping [43].

### 4. Device characterization
Fig. 4 shows the linear I-V characteristics of the graphene/Si Schottky junction for different temperatures ranging from 280 K to 315 K.

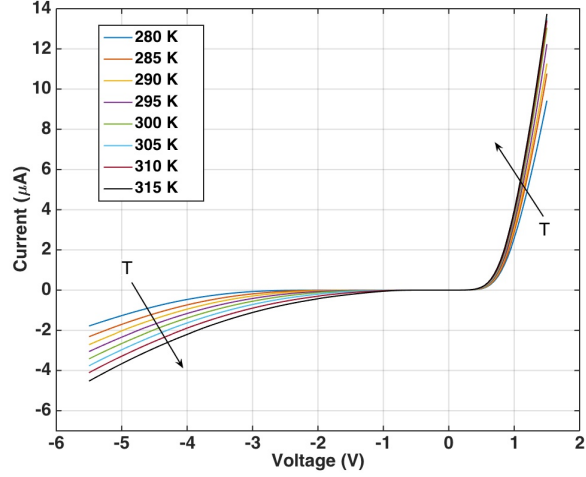

Figure 4. I-V characteristic of Graphene/Si junction for increasing temperature from 280 K to 315 K.

The rectifying I-V characteristic confirms the Schottky nature of the graphene/Si junction. Reverse (dark) current is about 8 nA at -1V, more than two order of magnitude lower than forward current of 3.5 µA at 1 V. The shape of the IV curve suggests fitting the experimental data with the following Eq. 2 [38]:

$$I = A_{PD} A^* T^2 e^{-\frac{q\phi_{B0}}{kT}} \cdot \left( e^{\frac{V - R_S I}{\eta k T}} - 1 \right) \quad (2)$$

where η is the ideality factor, k the Boltzman constant ($0.86266 \cdot 10^{-4}$ eV/K), T the absolute temperature, $R_s$ is the series resistance, $A_{PD}$ is the graphene area in contact with Si (part of the red dashed circle of Fig. 2(b) not overlapped to Au electrode) measured as 1127 µm² by optical microscopy, $A^*$ is the Richardson constant (32 A/cm²K² for p-type Si), $\Phi_{B0}$ is the graphene/Si Schottky barrier at zero bias. Indeed by using Eq. 2 to carry out the fitting process of the experimental data shown in Fig. 4 and taking $\Phi_{B0}$, η and $R_s$ as fitting parameters for any T, we get the following results:

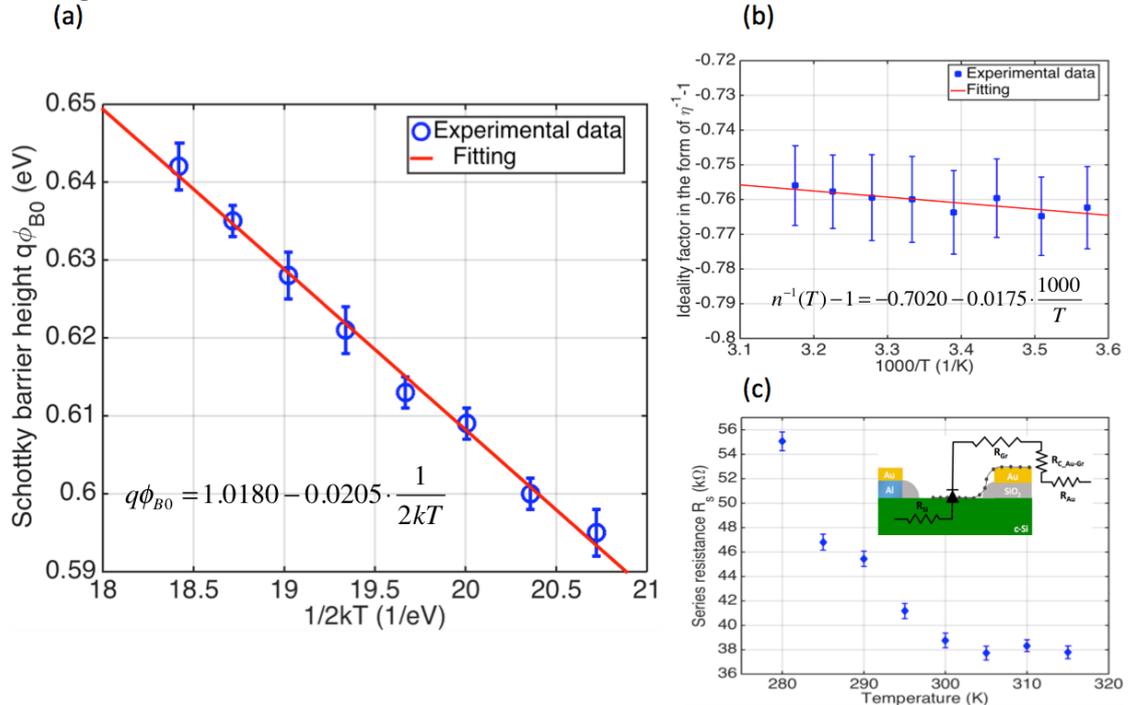

Figure 5. Results of the fitting operation carried out on the experimental data shown in Fig.4: (a) SBH $q\Phi_{B0}$ versus inverse temperature, (b) ideality factor in the form of $\eta^{-1}-1$ versus inverse temperature and (c) series resistance $R_s$ versus temperature (the inset shows all resistances in series to ideal graphene/Si diode).

Fig. 5(a) shows a linear dependence of the SBH on the temperature even if the standard thermionic emission (TE) theory would predict a temperature independent SBH. Temperature dependent SBH could be explained by taking into account both the tunneling contribution and the image force in the ideal TE theory. Indeed, in extrinsic semiconductors, carriers can tunnel across the Schottky barrier with a characteristic tunneling energy $E_{00}=(\hbar/2)\sqrt{N_a(m^*_h\varepsilon_{si})}$ where $N_a$ is the donor concentration, $m^*_h$ the hole effective mass, and $\varepsilon_{si}$ the dielectric permittivity of Si. In agreement with the Padovani's work [44], the tunnelling contribution can only be observed at thermal energy lower than characteristic energy $E_{00}$ (kT<<$E_{00}$), thus, in our case, being $N_a$~$10^{15}$cm$^{-3}$, $m^*_h$=0.55$m_0$ (where $m_0$=9.11x$10^{-31}$Kg is the electron mass) [38], and $\varepsilon_{si}$=11.4$\varepsilon_0$ (where $\varepsilon_0$ is the vacuum permittivity) a value $E_{00}$ of 0.23 meV can be achieved, much smaller than the thermal activation energy of ~27 and ~24meV at 315 and 280 K, respectively. This analysis shows that the tunnelling effect can't be responsible of the SBH dependence on the temperature. In addition, the model of image-force lowering can't be used because of the small $\Delta\Phi(T)$ SBH variation with the temperature in comparison with the experimental data. Indeed being $\Delta\Phi(T)=\sqrt{qE_{max}(T)/4\pi\varepsilon_{si}}$ where $E_{max}=\sqrt{2qN_aV_{bi}(T)/\varepsilon_{si}}$, $V_{bi}(T)=\Phi_{B0}-(kT/q)\ln N_v(T)/N_a$ and the Si density-of-state effective mass of the valence band $N_v=2(2\pi m^*_h kT/h^2)^{3/2}$, a value of only $q(\Delta\Phi(315K)-\Delta\Phi(280K))$=1.28x$10^{-4}$ eV, can be achieved despite of a variation of 0.047eV shown by experimental data reported in Fig. 5(a). A possible explanation about the dependence of the SBH on the temperature was given by Werner and Guttler: indeed, in a Schottky junction, the interface between the metal and semiconductor is not atomically flat but rough, with the result of spatial fluctuations of the built-in voltage $V_{bi}$ and consequently of inhomogeneities of the SBH. The authors proved that if a spatial Gaussian distribution of the SBH is assumed, the $\phi_{B0}$ measured by IV measurements follows the mathematical expression:

$$q\phi_{B0} = q\bar{\phi}_{B0} - \frac{q^2\sigma_{s0}^2}{2kT} \qquad (3)$$

where $\bar{\phi}_{B0}$ and $\sigma_{s0}$ are the mean SBH at T=0k, and the standard deviation of the Gaussian distribution. The $q\Phi_{B0}$ is plotted as a function of 1/2kT in the Fig. 5(a), yielding a mean SBH of $q\bar{\phi}_{B0} = 1.0180 \pm 0.0144$ eV and $q\sigma_{s0} = 0.1431 \pm 0.0026$ eV. It is worth mentioning that the current I across the interface depends sensitively on the detailed barrier distribution at the interface. Indeed, the spatial variations in the barrier causes the current I to flow preferentially through any SBH minimum, which is the value determined by IV measurements. It is worth noting that Fig. 5(a) shows a Schottky barrier of 0.621±0.004eV at 300K and zero bias in a good agreement with the theory where $q\Phi_{B0}=E_g-q(\Phi_{gr}-\chi_{Si})$=0.67eV, being the graphene workfunction $\Phi_{gr}$=4.5 eV [45], the Si electron affinity $\chi_{Si}$=4.05 eV [38] and the Si bandgap $E_g$=1.12 eV [38]. Despite of the good agreement with the SBH minimum, a large discrepancy with theory can be observed with respect to the measured mean SBH of 1.1018eV. This discrepancy could be linked to the presence of dangling bonds and charge traps at graphene/silicon interface responsible for Fermi level pinning [38]. The density of these traps can be evaluated by modelling the Schottky diode with an equivalent circuit shown in the inset of Fig. 6, taking into account the trap states. In Fig. 6 $C_j$, $C_{it}$ and $R_{it}$ are the junction capacitance, traps associated capacitance and resistance, respectively, where $\tau=C_{it}R_{it}$ can be defined as the trap lifetime while in the low frequency regime $D_{it}=C_{it}/q^2$ [46] is the trap density. The equivalent circuit shown in the inset of Fig. 6 can be converted into a frequency-dependent capacity $C_p$ in parallel with a frequency-dependent conductance $G_p$ given by:

$$\frac{G_p}{\omega} = \frac{C_{it}\omega\tau}{1+\omega^2\tau^2} \qquad (4)$$

$$C_p = C_j + \frac{C_{it}}{1+\omega^2\tau^2} \tag{5}$$

where ω is the angular frequency. The frequency-dependent conductance $G_p/\omega$ has been measured by a LCZ meter (Keithley, Model 3322) able to drive the sample with a known voltage sine wave signal at a certain frequency and to derive the impedance by precisely measuring the resultant current. Results are reported in Fig. 6. By performing a fitting operation between the experimental data and Eq. 4, the values of $C_{it}$, τ and $D_{it}$, can be easily derived. The same derivations could be obtained by Eq. 5, but in this case inaccuracy of extracted values is expected due to the sensitive dependence of $C_p$ to measured frequency ω.

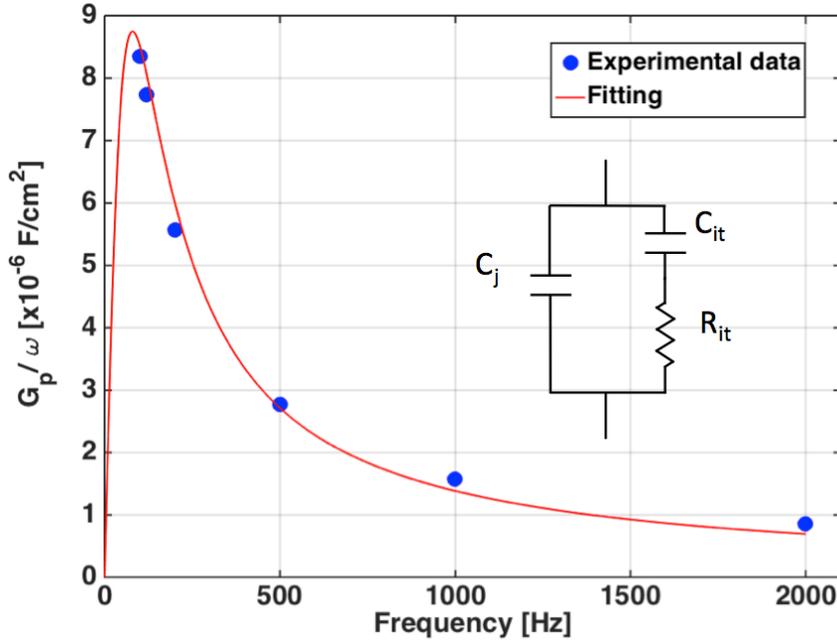

Figure 6. Conductance versus frequency plot. Inset shows equivalent circuit of the graphene/silicon diode including trap states.

The extracted $C_{it}$, τ and $D_{it}$ are $1.75 \times 10^{-5}$ F/cm², 2 ms and $1.1 \times 10^{14}$ eV$^{-1}$cm$^{-2}$, respectively. The extracted interface trap density is higher with respect to that one reported for classical SiO$_2$/Si interfaces [47].

Moving our attention on the ideality factor shown in Fig. 5(b), also its dependence on the temperature can be ascribed to spatial fluctuations of the built-in voltage $V_{bi}$ and written as: $n^{-1}(T)-1=-\rho_2+\frac{\rho_3 q}{2k}\cdot\frac{1}{T}$ [48] that predicts a general temperature dependence if both the mean barrier $\bar{\phi}_B(V)$ and the square of the standard deviation $\sigma_s^2(V)$ vary linearly with bias V according to $\bar{\phi}_B(V)-\bar{\phi}_{B0}=\rho_2\cdot V$ and $\sigma_s^2(V)-\sigma_{s0}^2=\rho_3\cdot V$ [48]. Thanks to the fitting process shown in Fig. 5(b), we find the coefficients $\rho_2$=0.702 and $\rho_3$=-3.018 mV for the graphene/Si Schottky junction. Being $\rho_2$>0 and $\rho_3$<0, we obtain that if bias V>0 increases, the value of the standard deviation $\sigma_s(V)$ decreases while the mean SBH $\bar{\phi}_B(V)$ increases in agreement with that one expected for a graphene/p-Si Schottky junction.

Finally, concerning series resistance, it can be viewed as the sum of many contributions: the resistance of the Au electrode $R_{Au}$, the contact resistance of the Au-graphene junction $R_{c\_Au-Gr}$, the graphene resistance $R_{Gr}$ and the bulk silicon resistance $R_{Si}$ as shown in the inset of Fig. 5(c). Among them, only the bulk Si resistance $R_{Si}$ could justify the exponential dependence on the temperature shown in Fig. 5(c), being the temperature dependence of $R_{Si}$ directly related to bulk resistivity ρ by the formula $R_{Si} \propto \rho = const\cdot e^{\frac{E_a}{kT}}$ [38] where $E_a$ is the activation energy.

Here we can try to speculate the cause of spatial fluctuations of the built-in voltage $V_{bi}$ (and consequently of SBH inhomogeneities), which could be ascribed to the carrier fluctuations on the graphene surface due to the graphene transferring process generating topographic corrugations and electron-density inhomogeneities as reported in Ref. [49].

Finally, the opto-electronic characterization is carried out using the experimental set-up of Fig. 7.

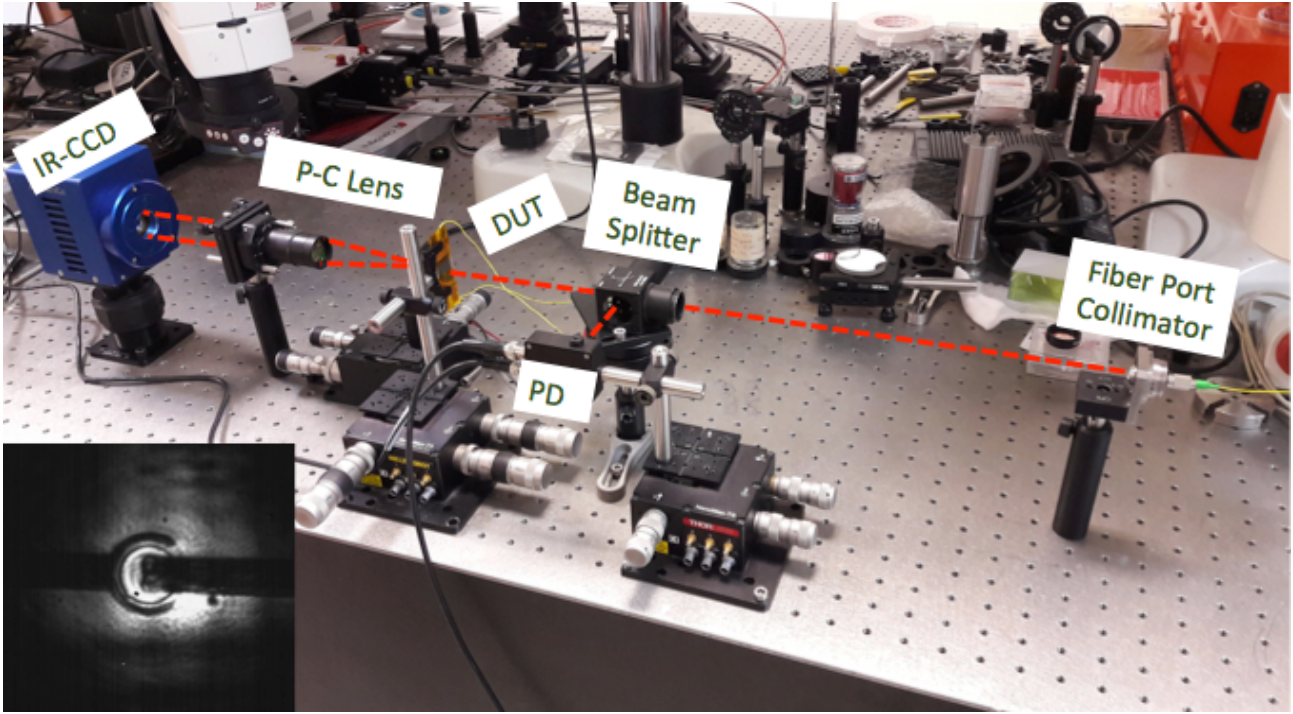

Figure 7. Experimental set-up for photocurrent measurements. In the inset MIR CCD image of the device.

A 2 μm light beam emitted by a diode laser (Thorlabs FPL2000S), has been collimated (beam diameter of 3 mm), chopped and split in two beams by a beam splitter. One beam is delivered onto the device under test (DUT). Transmitted light is collected by a 20X IR collecting objective microscope and addressed on a Mid-Infrared CCD in order to simplify the alignment procedure (see inset of Fig. 7). A lock-in amplifier measures the photocurrent produced by the DUT. Second beam is sent to a calibrated commercial InGaAs PD (Thorlabs PDA10D-EC) for incident optical power measurements. Finally, in order to calculate the internal responsivity, the measured incident optical power has been multiplied by the estimated SLG absorbance A to achieve the absorbed optical power. The SLG absorption has been calculated taking advantage of the Fresnel coefficient for thin films [50] in the Si/Graphene/Air three layer system. The SLG permittivity has been taken as:

$$\varepsilon = \varepsilon_\infty + i\frac{\alpha\lambda}{2d_{SLG}} \quad (6)$$

where $d_{SLG}$=0.335 nm the SLG's thickness and $\alpha = q^2/2\varepsilon_0 hc = 7.29 \times 10^{-3}$ is the fine structure constant (where q is the electron charge, $\varepsilon_0$ is the vacuum permittivity, h is the Plank constant and c is the vacuum speed) and $\varepsilon_\infty$=2.148 has been derived in Ref. [51]. The calculated SLG absorptance is shown in the up inset of Fig. 8 from which it is possible to extract a value of $A_{SLG}$=1.579 % at 2 micron. The resulting $I_{ph}$-$P_{abs}$ characteristic is reported in Fig. 8.

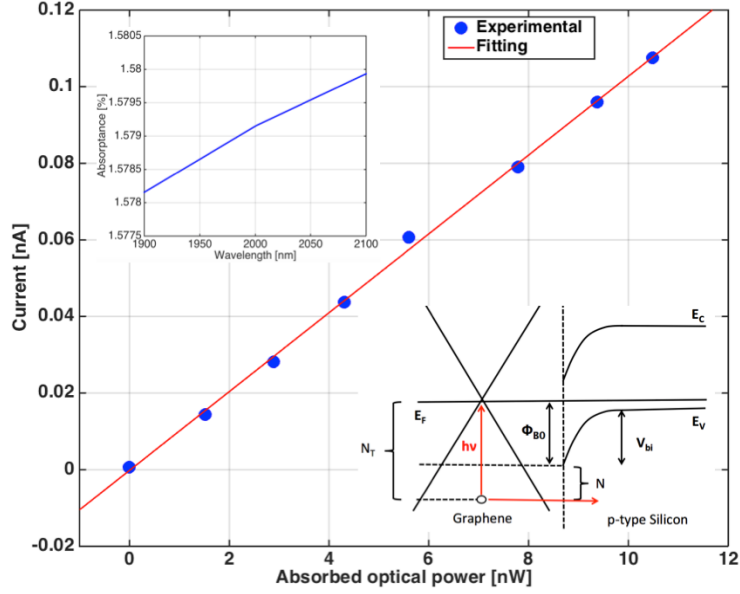

Figure 8. Measured photocurrent vs absorbed optical power and relative curve fit carried out by Eq. 7. Up inset: Calculated graphene absorptance versus wavelength. Down inset: Graphene/pSi Schottky junction energy band diagram.

As expected the $I_{ph}$-$P_{abs}$ behaviour is linear, no error bars are reported because measurements are very precise and accurate thanks to the lock-in technique. The slope is the internal responsivity that results 10.3 mA/W, corresponding to an external responsivity of 0.16 mA/W. It is worth noting that taking advantage of Eq. (1), the fitting operation with experimental data shown in Fig. 8 failed. Indeed both the quantum efficiency coefficient C and the SBH $\Phi_B$ were characterized by very large confidence intervals. In addition the SBH was calculated as $\Phi_B$=0.4588 eV in a very poor agreement with results obtained by the IV measurements. This is because Eq. 1 does not correctly describe IPE occurring through junctions involving graphene. Indeed, graphene is a two-dimensional material while the Elabd's theory (and so Eq. 1) was developed for junctions based on three-dimensional materials, i.e., metals. Thus, in our case the IPE theory should be further reviewed. By following the Elabd's approach [39], in the zero temperature approximation, the total number of possible excited states for holes is $N_T = \int_0^{h\nu} D(E) \cdot dE$ as shown in the down inset of Fig. 8, where $D(E) = \dfrac{2E}{\pi \hbar^2 v_F^2}$ [52] is the graphene density of state (DOS), E is the hole energy referred to the Fermi level, $\hbar$ the Plank constant, $v_F$ the Dirac velocity and $h\nu$ the energy of the incoming photons. On the other hand, the number of states N from which hole emission across the barrier $\Phi_B$ may occur is $N = \int_{\phi_B}^{h\nu} D(E) \cdot P(E) \cdot dE$ where P(E) is the emission probability. In three-dimensional material P(E) can be expressed as P(E)=(1-cos$\vartheta$) /2 [22] where $\vartheta$ defines the solid angle for the escape of hot holes with energy E [39]. P(E) changes in two dimensional material, indeed in graphene the π orbitals are always normal to graphene/Si interface, whereby hot hole momentum can only have two directions: one pointing toward Si and the other in the opposite versus. By following this line of reasoning it is possible to understand as the emission probability P(E) is not only independent on the energy E but also equals to ½ [26]. For this reason, the internal quantum efficiency can be written as: $\eta_i = \dfrac{N}{N_T} = \dfrac{1}{2} \cdot \dfrac{(h\nu)^2 - \phi_B^2}{(h\nu)^2}$. Finally, the resulting internal responsivity, i.e., $R_{int}$= $I_{ph}/P_{abs}$= $\eta_i$/h$\nu$ [A/W], offers a link between the photogenerated current and absorbed optical power that can be used as fitting equation for the experimental data shown in Fig. 8:

$$I_{ph} = \frac{1}{2} \cdot \frac{(h\nu)^2 - \phi_B^2}{(h\nu)^3} \cdot P_{ass} \quad (7)$$

Indeed, a SBH of $\Phi_B$=0.617 eV, in a really good agreement with results obtained by the electrical characterization, can be achieved. This result confirms also the validity of the proposed model summarized by Eq. 7. A simple derivation is that the IPE approach is able to provide the minimum value of SBH because the photo-excited carriers flow preferentially through the barrier minima. It is worth mentioning that device responsivity is mainly limited by the low graphene absorbance (only 1.579%) whose value could be further increased by taking advantage of the optical field enhancement occurring inside Fabry-Perot optical microcavities [6,53]. Our measurements prove that by combining graphene with IPE, Si-based free-space PDs operating at 2 μm can be developed for FSO, OCT and LIDAR applications.

## 5. Conclusions

We demonstrated a free-space illuminated SLG/Si Schottky PD at 2 micron. The photodetection mechanism is based on internal photoemission at the SLG/Si interface. The graphene/silicon junction has been electrically characterized at temperatures ranging from 280 to 315 K showing a linear dependence of both the SBH and ideality factor on the temperature that has been ascribed to spatial fluctuations of the built-in voltage $V_{bi}$ and consequently of inhomogeneities of the SBH due to the transferring process of graphene. In addition the quality of the graphene/Si interface has been investigated measuring an interfacial trap density of 1.1 x $10^{14}$ $eV^{-1}cm^{-2}$. Finally, we showed photoresponse with internal (external) responsivity of about 10.3 mA/W (0.16 mA/W) that can be only predicted by modifying the IPE theory commonly used for junctions involving three-dimesional materials.

Responsivity is limited by the low SLG absorption and it could be increased by taking advantage of high-finesse Fabry-Perot microcavity able to provide enhanced SLG absorption. Our devices pave the way for developing hybrid graphene-Si free-space illuminated PDs operating at 2 μm, for FSO communications, OCT and LIDAR applications.